\documentclass[preprint2]{aastex63}

\usepackage{float}
\usepackage{gensymb}
\usepackage{amsmath}
\usepackage{dblfloatfix}
\usepackage{savesym}
\savesymbol{tablenum}
\usepackage{dirtytalk}
\usepackage{siunitx}
\restoresymbol{SIX}{tablenum}
\usepackage{graphicx}
\turnoffedits

\shorttitle{DAXSS QUIESCENT SOLAR SOFT X-RAY SPECTRA}
\shortauthors{Schwab and Sewell et al.}
\graphicspath{{./}{figures/}}
\begin{document}

\title{Soft X-Ray Observations of Quiescent Solar Active Regions using Novel Dual-zone Aperture X-ray Solar Spectrometer (DAXSS)}

\author[0000-0002-1426-6913]{Bennet D. Schwab}\altaffiliation{These authors contributed equally to this work.}
\affiliation{Ann \& H.J. Smead Department of Aerospace Engineering Sciences, University of Colorado at Boulder, 3775 Discovery Dr., Boulder, CO 80303}

\author[0000-0002-3440-2492]{Robert H. A. Sewell}\altaffiliation{These authors contributed equally to this work.}
\affiliation{Laboratory for Atmospheric and Space Physics, University of Colorado at Boulder, 3665 Discovery Dr., Boulder, CO 80303}

\author[0000-0002-2308-6797]{Thomas N. Woods}
\affiliation{Laboratory for Atmospheric and Space Physics, University of Colorado at Boulder, 3665 Discovery Dr., Boulder, CO 80303}

\author[0000-0001-8702-8273]{Amir Caspi}
\affiliation{Southwest Research Institute, 1050 Walnut St Suite 300, Boulder, CO 80302}

\author[0000-0002-3783-5509]{James Paul Mason}
\affiliation{Laboratory for Atmospheric and Space Physics, University of Colorado at Boulder, 3665 Discovery Dr., Boulder, CO 80303}

\author[0000-0002-4103-6101]{Christopher Moore}
\affiliation{Harvard-Smithsonian Center for Astrophysics, 60 Garden Street, MS 58, Cambridge, MA 02138, USA}

\begin{abstract}

The Dual-zone Aperture X-ray Solar Spectrometer (DAXSS) was flown on 2018 June 18 on the NASA 36.336 sounding rocket flight and obtained the highest resolution to date for solar soft X-ray (SXR) spectra over a broad energy range. \edit1{This observation was during a time with quiescent (non-flaring) small active regions on the solar disk and when the 10.7 cm radio flux (F10.7) was 75 solar flux units (1 sfu = 10--22~W/m$^2$/Hz).} The DAXSS instrument consists of a LASP-developed dual-zone aperture and a commercial X-ray spectrometer from Amptek that measures solar full-disk irradiance from 0.5--20~keV with a resolving power of 20 near 1~keV. This paper discusses the novel design of the spectrometer and the instrument characterization techniques. Additionally, the solar measurements obtained from the 2018 sounding rocket flight are analyzed using CHIANTI spectral models to fit the temperatures, emission measures, and relative elemental abundances of the solar corona plasma. The abundance of iron was found to be \edit1{35} percent higher than expected in the quiescent sun's corona suggesting either that our spectral models require additional sophistication or that the underlying atomic database may require updates. Future long-term systematic observations of this spectral range are needed. DAXSS will fly on the INSPIRESat-1 CubeSat in late-2020, and its SXR spectral data could provide further insight into the sources of coronal heating through modeling the changes of relative elemental abundances during developments of active regions and solar flaring events.
\end{abstract}



\section{Introduction} \label{sec:intro}
The solar corona, during quiescent,  non-flaring periods, has very hot temperatures of higher than ~1 MK and is about $100$ times hotter than the Sun's inner layers, the chromosphere and photosphere \citep{golub2010solar}. The source of the corona's much higher temperature is not yet fully understood and remains one of the fundamental unanswered questions in solar physics \citep{klimchuk_2006}. There are two main theories that aim to explain the heating process. The first is through magnetic reconnection of field lines in the corona that cause "nano-flares" that can heat coronal plasma to temperatures of $10-15\ MK$ in family with solar flares \citep{parker_1988}. The second is dissipation of Alfv\'en waves that heats the plasma to relatively narrow distributions of coronal temperatures of $1-3\ MK$ \citep{asgari-targhi_2012}.

\edit1{Solar X-ray (SXR) and hard X-ray (HXR) observations are important for understanding this heating problem because those wavelength ranges include continuumm and emission lines from the hot corona \citep{fletcher_2011} that can reveal the sources of coronal heating.} The two dominant processes that contribute to the continuum are free-free Bremsstrahlung emission and free-bound recombination emission. The slope of the continuum in this SXR range is highly \edit1{sensitive} to the temperature(s) of the emitting plasma. The normalization of the continuum is then an indicator of the emission measure. Some of the less intense emission lines in this energy range are not distinguishable above the intensity of the continuum; however, some of the brighter emission lines are identifiable and may be used to model their relative elemental abundances.

A key diagnostic for exploring the sources of coronal heating is looking at the change in abundance for \edit1{low first-ionization potential (FIP)} elements, i.e., elements with FIP below about 10~eV, such as Si, Ca, and Fe. The abundance change relative to the photospheric abundance is expected to be about $2-4$ for coronal closed magnetic field features and closer to 1 (photospheric) for open field features \citep{laming_2015}. Furthermore, heating due to magnetic reconnection could show lower elemental abundances in the corona than that from Alfv\'en dissipation heating due to plasma injection into the coronal loops from the chromosphere \citep{warren_2014}. Therefore, analyzing the coronal abundance \edit1{could be} an effective diagnostic technique to identify that both forms of heating are important in the corona and by how much.

One way to gain insight into which heating process may be dominant on the sun during different activity levels is to analyze the elemental abundance for several emission lines in the SXR regime. The range of energies between 0.5--10~keV contains many emission lines of the hot plasma in the corona. For solar observational instruments there has been a spectral gap in the SXR range of 0.2~keV and 3~keV, between the usable ranges of the Solar Dynamics Observatory (SDO) \citep{sdo_2011} Extreme ultraviolet Variability Experiment (EVE) \citep{woods_2012} and the Reuven Ramaty High Energy Solar Spectroscopic Imager (RHESSI) \citep{rhessi_2002} satellites \citep{smith_2003, woods_2012}, \edit1{as discussed by \citet{mason_2019}}. Some energies within this gap have never been measured in any resolution and are also outside of the \edit1{energy range measured by the Dual-zone Aperture X-ray Solar Spectrometer (DAXSS) (which is analyzed in this paper)}, while other energies have been measured only coarsely except for recent sporadic measurements in the past 20 years. In addition the measurements from these observatories are limited, as RHESSI is designed and most sensitive for flare X-rays and is only marginally sensitive to quiescent SXR emissions \citep{mctiernan2009} and there are very few coronal EUV lines in the 5-10 MK range essential to probe active region heating \citep{caspi2015}. Furthermore, other instruments such as Yohkoh/BCS and CORONAS-F/RESIK have narrow passbands of about 0.05 keV \citep{mariska_2006} in the SXR spectrum and do not have much information about the underlying continuum. \edit1{Solar SXR spectra can be derived through differential emission measure (DEM) diagnostics from a set of EUV and SXR images \citep{Su_2018}} but direct spectrometry is the best method for studying the nature of the SXR and are needed in order to validate these derived measurements . 

The University of Colorado Boulder (CU) efforts to close the SXR measurement gap includes using wide passband spectrometers aboard sounding rocket flights (2012, 2013), the \edit1{Miniature X-ray Solar Spectrometer CubeSat  (MinXSS-1)} (May 2016 -- May 2017), the MinXSS-2 CubeSat that launched on 2018 December 3, and the \edit1{Dual-zone Aperture X-ray Solar Spectrometer (DAXSS)} on NASA 36.336 sounding rocket flight  \citep{caspi2015,mason_2016,woods2017,moore2018,mason_2019}. The Mercury MESSENGER mission has the Solar Assembly for X-ray (SAX) sensor with similar SXR spectral measurements but at lower energy resolution than MinXSS or DAXSS \citep{sax2015} \edit1{see Table \ref{table:res_comp}}.

One goal for the MinXSS-1, MinXSS-2, and DAXSS missions is to aid in finding a solution to coronal heating processes, plasma temperature, and composition change for active regions during the different solar cycle phases. This goal is addressed by modeling the SXR spectra to determine the plasma temperature, density, and composition. \edit1{This is a method which has been done in many different ways.} One approach, referred to as the \edit1{two-temperature (2T)} model, is to fit the spectra with a hot temperature component and a cooler temperature component simultaneously with a singular \edit1{abundance factor (AF)} \citep{caspi2015,woods2017}. \edit1{This AF parameter is a constant that is multiplied by the Feldman standard extended coronal (FSEC) abundance values \citep{feldman1992,Landi2002}}. Another approach is to do multiple temperature derivation by estimating a DEM distribution over temperature and then adjusting the DEM function until the model fits the measured solar spectrum \citep{caspi2015}. In both approaches, the CHIANTI atomic database \citep{dere_1997,dere_2019} can be used to specify the emission brightness based on the plasma parameters of temperature, emission measure, and abundance.

The DAXSS rocket flight on 2018 June 18 provides a sample of the quiescent-sun, or non-flaring, activity that can be combined with other instrumental observations. DAXSS has two advantages over MinXSS for making quiescent measurements: DAXSS has improved energy resolution and its sensitivity is higher over a wider energy range. Analysis of this rocket measurement, \edit1{along with additional similar measurements and analyses in the future}, allows for the formulation of temperatures, emission measures, and relative abundances over different activity levels over the solar cycle. 

\edit1{This paper discusses the novel design of the DAXSS instrument and its improvements from earlier flights of similar spectrometers (Sec. ~\ref{sec:design}). The instrument calibration process is also described including gain and offset calibration, spectral resolution characterization, linearity of response assessment, field of view sensitivity, and detector responsivity (Sec.~\ref{sec:cal}). In Section~\ref{sec:model} we present the DAXSS observations from the NASA 36.336 sounding rocket flight and our solar spectral modeling of this data. In Section~\ref{sec:compare} the measurements, models, and results from this paper are compared with previous SXR measurements, DEM models, and GOES \citep{goes2020} measurements from the same time period. Finally, in Section~\ref{sec:conclusion} there is a discussion of our results and the future of DAXSS.}

\section{Dual-zone Aperture Design}\label{sec:design}
The Amptek X-123 FAST SDD X-ray spectrometer, \edit1{used on DAXSS}, includes a Si drift detector in a vacuum housing with a beryllium \edit1{filter} window. The Si depletion depth of 500~$\mu$m and a Be window thickness of 12.5~$\mu$m provides a sensitivity to X-rays from $\sim$0.5~keV to $\gtrsim$20~keV. The SDD has a two-stage thermoelectric cooler which keeps measurement noise low and Amptek has enhanced the X-123 electronics to improve energy resolution to $\sim$0.07~keV FWHM at 1~keV compared to about $\sim$0.20~keV FWHM at 1~keV for the earlier generations of the X-123 detectors flown on previous SDO/EVE rocket flights and the MinXSS satellites \citep{moore2018}.

\begin{figure}[H]
    \centering
    \includegraphics[width=\linewidth]{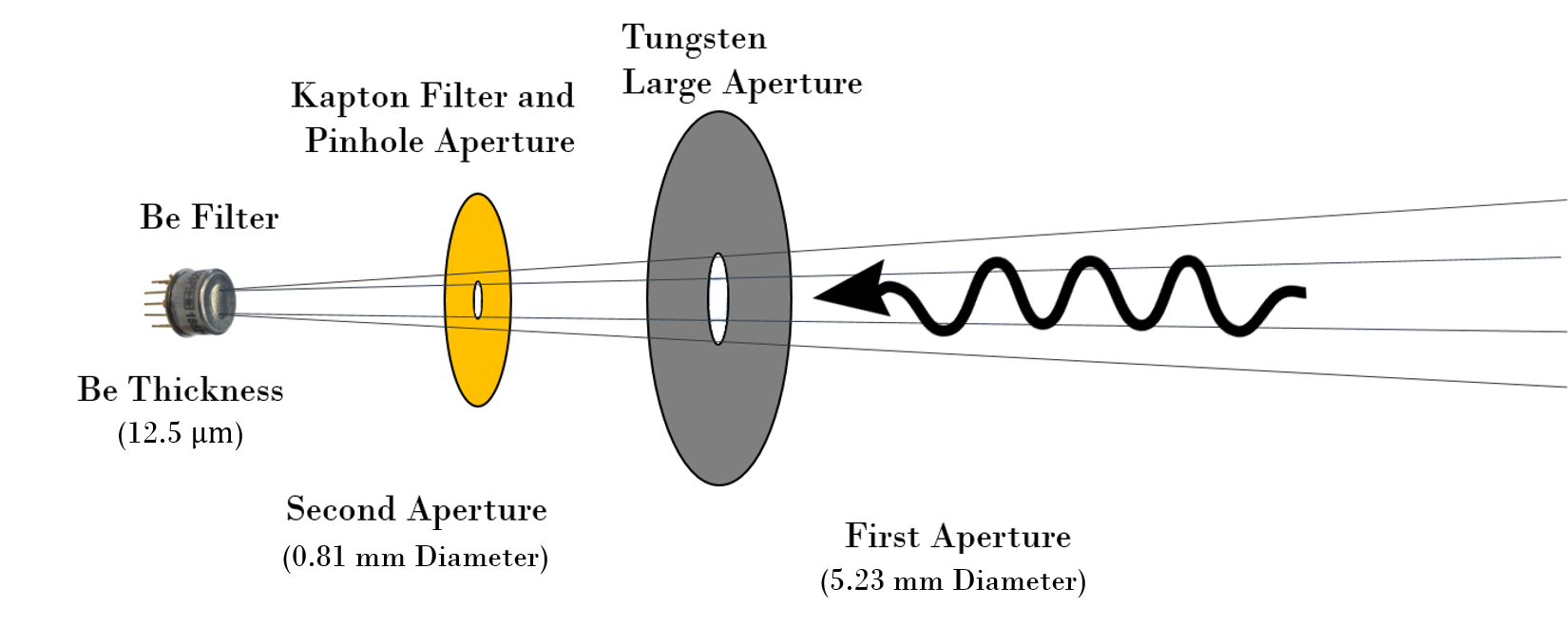}
    \caption{The concept for the dual-zone aperture design with the tungsten and Kapton apertures of differing diameters.\label{fig:cartoon}}
\end{figure}

In addition to the improved X-123 detector, \edit1{DAXSS also features a modified aperture designed compared to the previous flights of the instrument} (Figure~\ref{fig:cartoon}). The diameter of the larger primary aperture was widened and a Kapton filter was placed behind it. Widening the primary aperture allows for more photons of all energies to enter the detector's field of view, in particular the comparatively rarer high energy photons. The Kapton filter is added to attenuate the more prevalent lower energy photons as to not saturate the detector. A small pinhole aperture is laser etched into the center of the Kapton allowing some light to pass through unaffected and thus not attenuating all lower energy photons.

\begin{figure}[H]
    \centering
    \includegraphics[width=.8\linewidth]{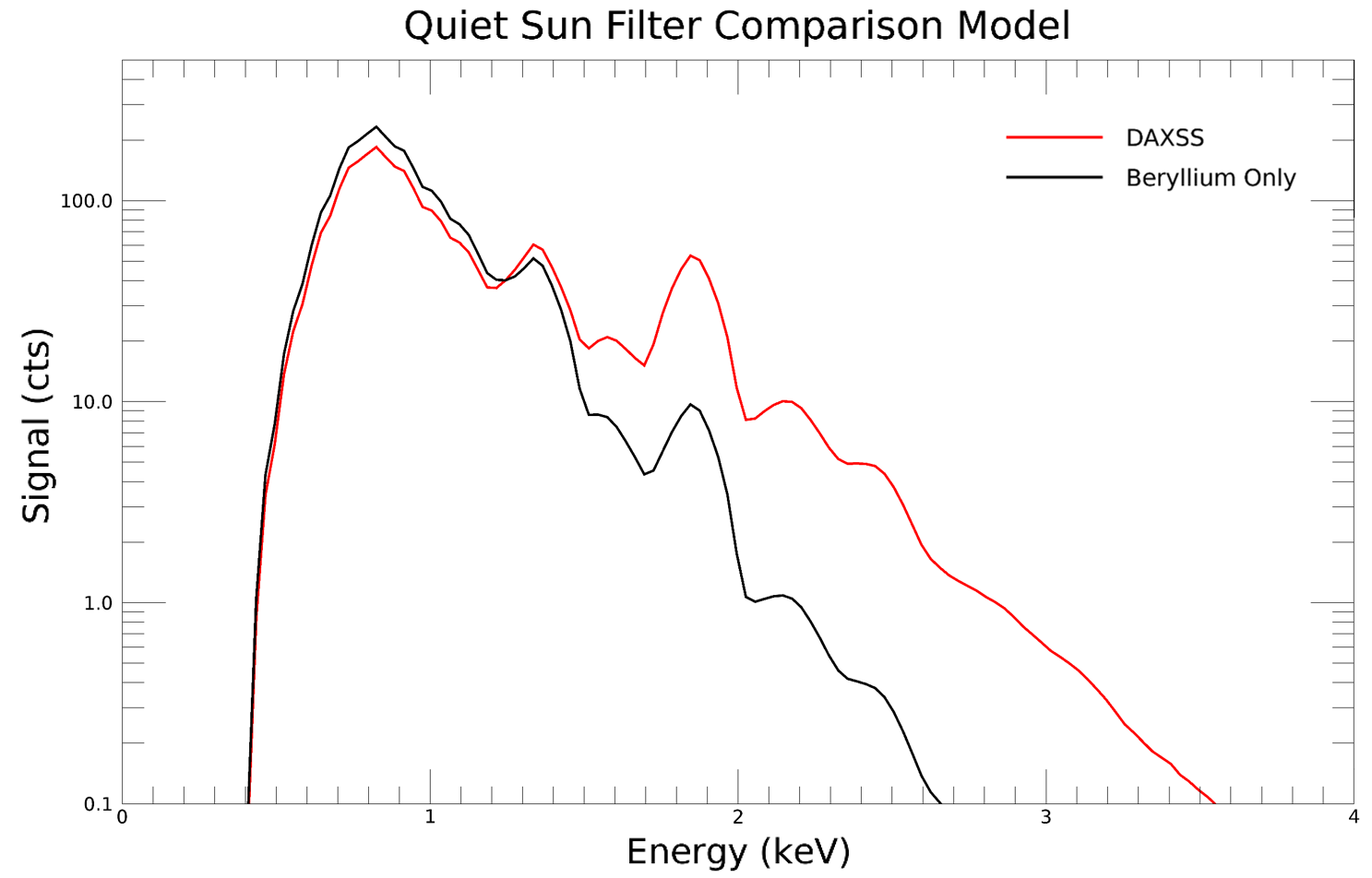}
    \caption{The modeled theoretical difference between the Amptek X-123 standalone detector and the X-123 with the dual-zone aperture design under quiet solar conditions with the same input flux and equal total observed signal. The DAXSS aperture enhances the higher energy photon measurements while only slightly limiting the lower energy measurements where the spectrum is already maximized. \label{fig:filter_comp}}
\end{figure}

The intended result of this new aperture design is to detect more photon events in the 1.5--20~keV spectrum without saturating the Si diode with lower energy events as shown in Figure~\ref{fig:filter_comp}. A model of the detector response was calculated using Henke transmission coefficients \citep{Henke1993XRayIP}. This response was used along with the \edit1{CHIANTI v7.0 \citep{dere_1997} DEM for quiet-sun to model theoretical SXR spectra} to determine the appropriate large and small aperture sizes and Kapton filter thickness for the detector given the expected phase in the solar cycle for the 2018 June rocket launch. \edit1{From this analysis  the tungsten aperture was chosen to be 5.232~mm with a field of view (FOV) of $\pm$ 4\degree, and the pinhole diameter in the Kapton was chosen to be 0.813~mm with a Kapton sheet thickness 25~$\mu$m.}

\section{Instrument Calibrations}\label{sec:cal}
The electronic settings and spectral resolution of DAXSS were determined and characterized using radioactive (gamma) sources and X-ray scattering targets. The linearity and spectral response were determined from measurements at the National Institute for Standards and Technology (NIST) Synchrotron Ultraviolet Radiation Facility (SURF) \citep{surf_2011}.

\subsection{Gain and Offset Calibration}\label{subsec:gain}
To take meaningful spectral measurements with DAXSS the detector bin to incident photon energy correlation was determined. This was done using the Amptek Mini-X X-ray source to fluoresce known materials and measure the emission lines with the X-123 detector as well as measure the emission lines from a radioactive source of Fe-55. As these line measurements are at well known energies we can determine a linear fit, \edit1{with a slope and an offset parameter}, relating detector bin to observed photon energy. Once this relationship was determined, the detector gain was adjusted so that the instrument is sensitive to energies from 0-20~keV over its 1024 channels (energy bins). 

\subsection{Spectral Resolution}\label{subsec:spectral_res}
\begin{figure}[H]
    \centering
    \includegraphics[width=\linewidth]{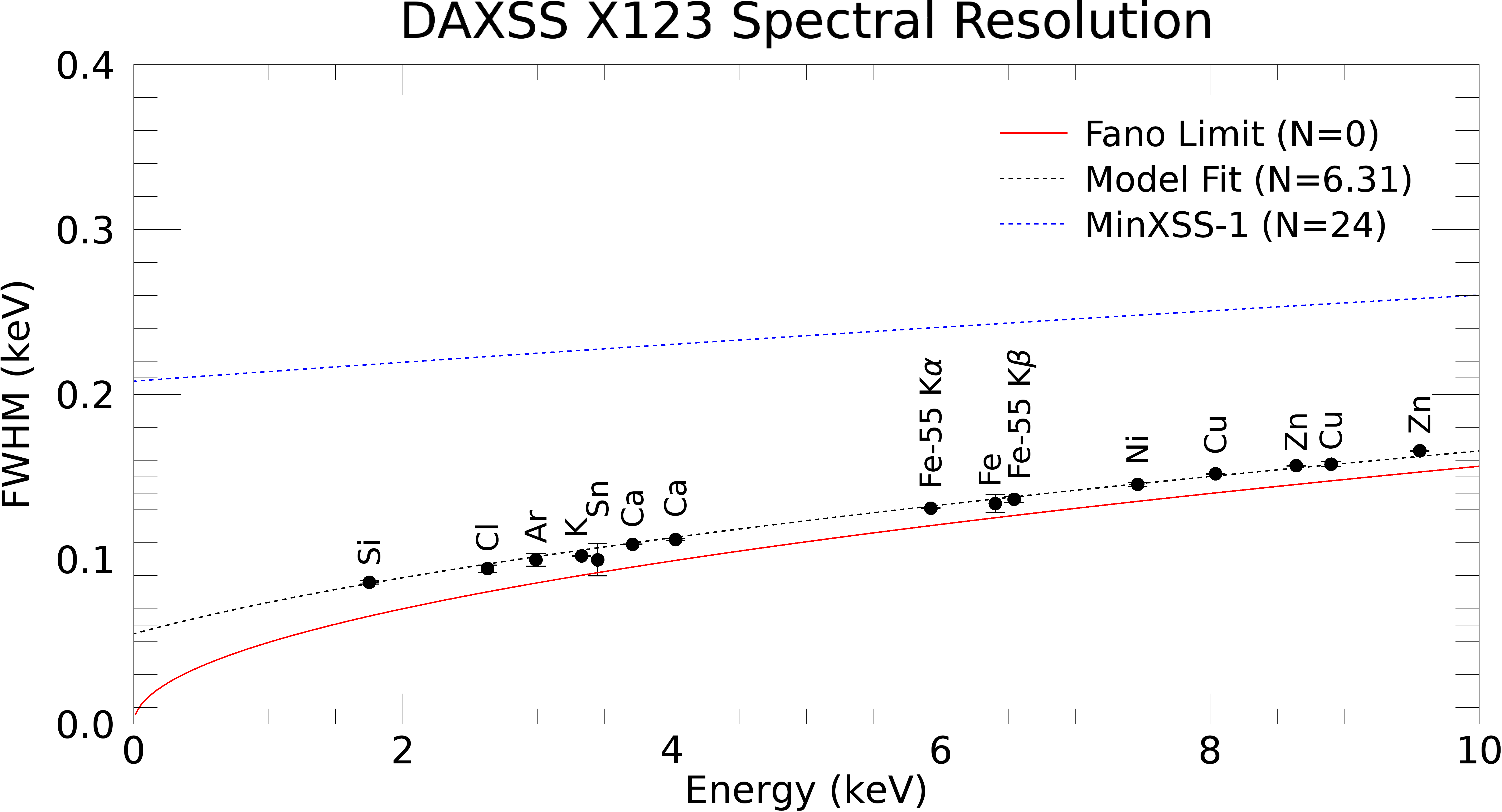}
    \caption{\label{fig:res}FWHM measurements of X-ray fluorescence from several known sources and the decay products of $^{55}$Fe. The FWHM of each line was determined by fitting a Gaussian to each emission line and taking the full width of the Gaussian at half of the Gaussian fit's maximum. The red line indicates the theoretical Fano Noise limit for Si and the dashed black and blue lines are the model fit for the resolution of the DAXSS X-123 FAST SDD detector and the MinXSS-1 X-123, respectively, using Equation~\ref{eqn:fano}.}
\end{figure}

The spectral resolution of DAXSS was characterized by using the same emission lines as were used for the gain and offset calibration, described in subsection~\ref{subsec:gain}, and measuring the full width at half maximum (FWHM) of the characteristic X-ray line emission signal seen by the X-123 detector.  The FWHM of the measured spectral emission and decay lines characterize the resolution of the detector at the known emission energies. The resolving power of X-ray emissions for Si based detectors are limited by Fano noise \citep{knoll_2000}, which is the intrinsic statistical variation in the electron-hole pair generation per event. The spectral resolution of the detector can be modeled as the Fano noise for Si with an additional term for other systematic contributions to the noise as described by 
\begin{equation}\label{eqn:fano}
    \textrm{FWHM}=2.35\omega\sqrt{F\frac{E_{ph}}{\omega}+N^2}
\end{equation}
where the factor of 2.35 is due to the relation between the Gaussian standard deviation and the FWHM, $F$ is the Fano factor of the material ($\sim$0.12 for Si), $E_{ph}$ is the energy of the photon emission \edit1{(in eV)}, $\omega$ is the average excitation potential for the material ($\sim$3.68~eV for Si at the detector operating temperature of 225~K), and $N$ is the systematic noise of the detector. \edit1{Using the FWHM from measured emission lines and Eqn.~\ref{eqn:fano} a best fit value for N was determined to be 6.45 (shown in Figure~\ref{fig:res}), which allows us to apply Eqn.\ref{eqn:fano} to get the spectral resolution across the entire detector sensitivity range of 0--20~keV.} The DAXSS energy resolution at 1 keV is a factor three better than the MinXSS X-123 energy resolution.
\subsection{Linearity of Response}
\begin{figure}[H]
    \centering
    \includegraphics[width=\linewidth]{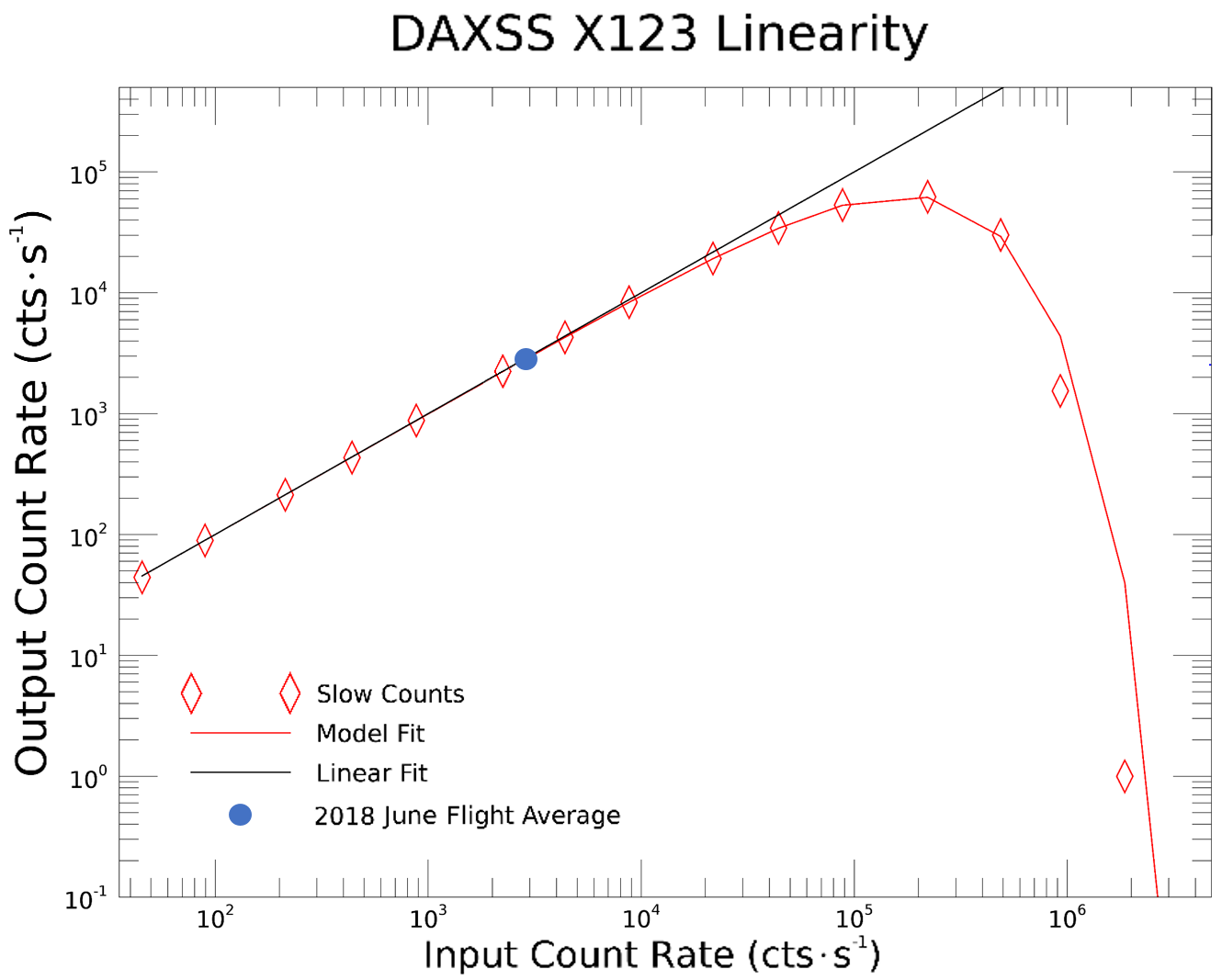}
    \caption{\label{fig:linearity} DAXSS detected \edit1{slow count rate ($C_s$) from actual input  count rate ($C_{in}$)} with 1.2~$\mu$s peaking time. The black line is the linear relationship between the dead time corrected input count rate and the output spectrum count rate. The red line indicates the dead time model fit ($C_{model}$ with $\tau_{ds}=2.875~\mu$s) to the measured output count rate ($C_s$). The blue dot is the average output count rate seen during the 2018 June rocket flight, given the observed input count rate, which is well below the region in which dead time effects become significant.}
\end{figure}

Being a photon-counting detector, it is important to understand DAXSS's linearity of response with light source intensity. The linearity of the detector was assessed at NIST SURF (Figure~\ref{fig:linearity}). Here, all the discussed count rates are the total measured counts across all energy bins during an integration period divided by the time of the integration. Looking at the relationship between the measured count rate and the actual rate provides insight into the severity of dead-time and pile-up effects \edit1{\citep{knoll_2000}}, the effects of which are described below.

The X-123 FAST SDD has two primary counting channels. The slow counter channel is used to create the X-123 spectrum and has a tunable peaking time (1.2~$\mu$s for DAXSS). The fast counter has a shorter peaking time of 100~ns and an effective pair resolving time of $\sim$120~ns ($\tau_{pair}$, $\tau_{df}$). Because of these features the fast channel operates in anti-coincidence mode with the slow channel. In this way photon peak pile-up events, in which more than one photon is absorbed within the peaking time and recorded as one photon with the sum of the all the photon energies, is minimized but may still occur \edit1{\citep{dp5_high_rates}}. 

Dead time refers to the time after each measured event during which the instrument cannot record another event. For high 'true' count rates, losses due to dead time can be significant but the input count rate can still be approximated \edit1{with the he dead-time correction for the fast counter channel model given by} \citep{redus2008, moore2018}:
\begin{equation}\label{eqn:dead_fast}
    C_{in}=\frac{C_f}{(1-C_f\tau_{df})}
\end{equation}
where $C_{in}$ represents the 'true' input count rate, $C_f$ is the count rate measured by the fast counter and $\tau_{df}$ is the pair resolving time of the fast counter. From this, the corresponding count rates of the slow counter that are used to generate the spectrum are given by the model:
\begin{equation}\label{eqn:dead_model}
    C_{model}=C_{in}e^{-C_{in}\tau_{ds}/2}
\end{equation}
where $\tau_{ds}$ is the slow counter dead time. We can see in Figure~\ref{fig:linearity} that this model agrees reasonably well with the NIST SURF \edit1{slow count data ($C_s$)} with the slow counter dead time found to be $\tau_{ds}=2.875~\mu$s. Thus, with the models in Equations~\ref{eqn:dead_fast} and \ref{eqn:dead_model} the true count rate during a DAXSS observation can be deduced up to 100,000 counts per second. 

\subsection{Field of View (FOV) Sensitivity}
\edit1{As DAXSS is a single-pixel detector it} integrates the incident radiation across its full FOV to create its measured signal. So while DAXSS does not have spatial resolution, its sensitivity does vary slightly with angle of incidence of the incoming solar radiation.  The DAXSS front aperture size and distance from the sensor defines a $\pm4\degree$ \edit1{($8\degree$ total)} clear field of view (FOV). The FOV map obtained from SURF Calibrations indicate a $<$0.5\% variance over the $\pm1\degree$ central FOV. Pointing during the 2018 June rocket flight was maintained well within this central FOV while spectral measurements were being taken.

\subsection{Detector Responsivity}\label{detector_responsivity}
The DAXSS instrument spectral efficiency was determined through a series of measurements conducted at NIST SURF. The absolute synchrotron spectral irradiance is used to determine the detection efficiency of the DAXSS instrument. The X-123 count rate \textit{per} energy bin, \edit1{denoted as j, is ($C_{bin,j}$), in units of counts$\cdot$ s$^-1$}, can be calculated by Equation~\ref{eqn:surf_counts_1} and Equation~\ref{eqn:surf_counts_2} \citep{moore2018}:
\begin{equation}\label{eqn:surf_counts_1}
    C_{bin,j}=\int_{E_{min},j}^{E_{max},j}\left[\Gamma(E_{det})\right]dE_{det}
\end{equation}
\begin{multline}
    \label{eqn:surf_counts_2}
    \Gamma(E_{det})=\int_0^\infty\int^{\Omega_\odot}S(E_{ph},\Omega)\cdot\\ \left[R_{X123}(E_{ph},\Omega,E_{det})\right]d\Omega dE_{ph}
\end{multline}
where $S(E_{ph},\Omega)$ is the incoming X-ray signal---depending on the photon energy $E_{ph}$ and the solid angle $\Omega$---and $R_{X123}(E_{ph},\Omega,E_{det})$ is the detected energy bin redistribution function. The energy bin redistribution function maps photon events to detected energy counts which depends on the geometric area and transmission of the each of the two concentric apertures, the detector resolution (Figure~\ref{fig:res}) and the FOV sensitivity. This mapping can be inverted to take detected counts and create an estimate of the incident photon flux which will be discussed more in Section~\ref{sec:model}. 

\edit1{In reality this photon-count redistribution function would map all potential incident photon energies to all possible deposited detector energy bins. Thus, one can interpret this redistribution function as a detector response matrix (DRM) with columns that connect incident photon energies to the rows, X-123 energy bins, in which they are deposited. However, because the leading contribution to the detector responsivity is due to the transmission efficiency of the two filters and the Si photopeak response, the redistribution function can be described by a detector response array (DRA) in the form}
\begin{multline}
    \label{eqn:resp}
    R_{X123, energy\ bin}=[T_{Be}(E_{ph})A_{small}+\\T_{both}(A_{large}-A_{small})]R_{Si}(E_ph)G(E_{det},E_{ph})
\end{multline}
  where $T_{Be}$ and $T_{Both}$ are the transmission of beryllium and of both beryllium and Kapton, respectively---calculated using Henke transmission coefficients \citep{Henke1993XRayIP}, and $A_{small}$ and $A_{large}$ are the geometric areas for the small aperture in the Kapton filter and large aperture in the tungsten cover, respectively. $R_{Si}$ is the responsivity of the X-123 Si diode.  $G(E_{det},E_{ph})$ is the redistribution of photon energy to measured energy which can be described as a Gaussian with FWHM from Equation~\ref{eqn:fano}. 
  
\begin{figure}[H]
    \centering
    \includegraphics[width=\linewidth]{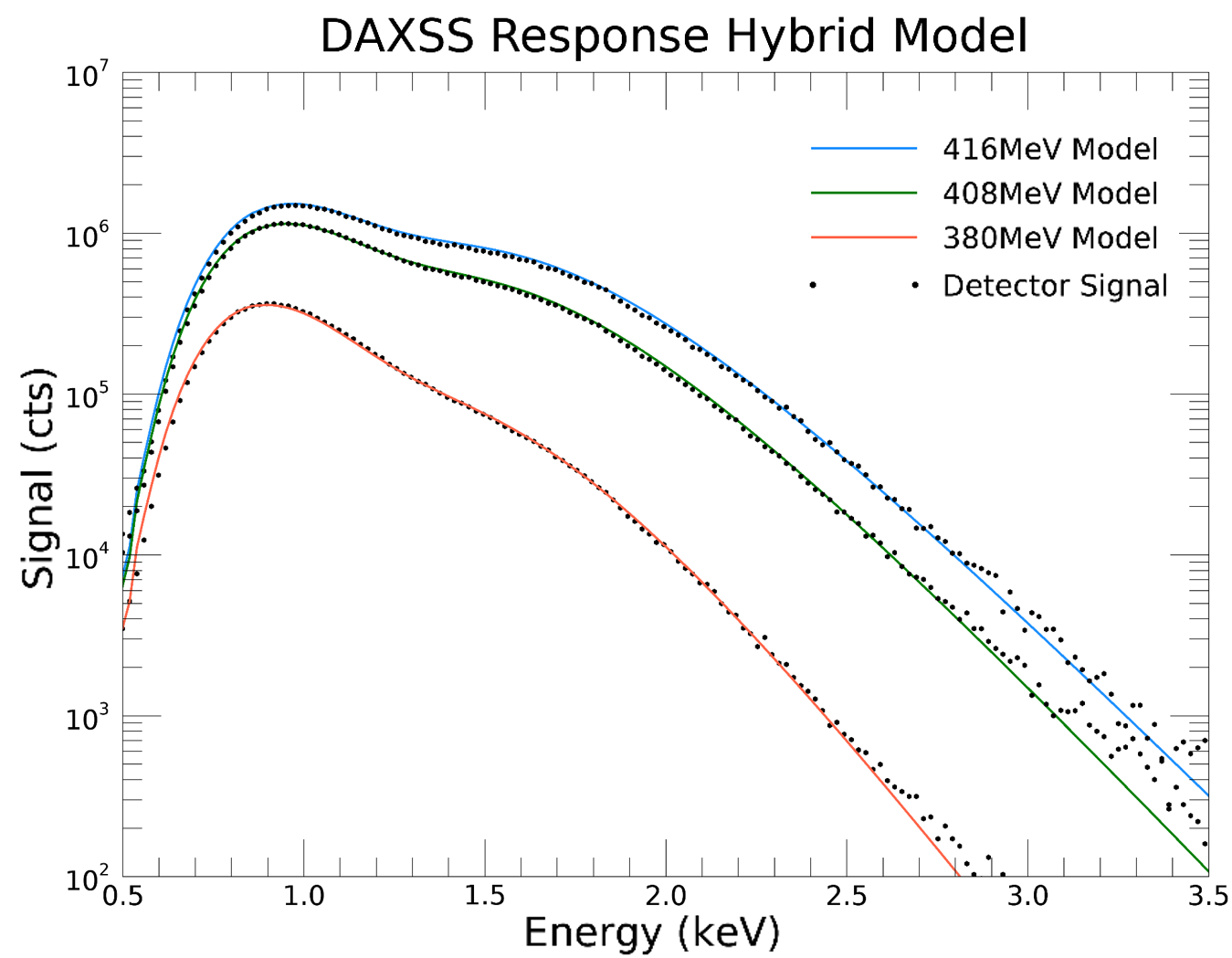}
    \caption{\label{fig:surf_model}NIST SURF X-ray spectrum measurements from the DAXSS instrument along with model fit described by Equations~\ref{eqn:surf_counts_1}-\ref{eqn:resp}.}
\end{figure}

 With the NIST SURF spectra, the relationships described in Equations~\ref{eqn:surf_counts_1}-\ref{eqn:resp}, and the spectral resolution fit found in Section~\ref{subsec:spectral_res} forward modeling can be done to find a best fit for the only unknowns in our DRA: the sensor Si photopeak response and the filter transmissions of Be and Kapton. The sensor Si thickness, however, affects the energy response more for energies $>$10~keV and did not need to be fitted for the SURF data and as such was assumed to be the thickness provided by Amptek (500~$\mu$m). The SURF synchrotron beam energy is adjustable so that one can calibrate over different energy ranges, with beam energies between 380 MeV and 416 MeV being best for DAXSS calibrations. Fitting this model in Equation~\ref{eqn:surf_counts_1} to this SURF data, shown in Figure~\ref{fig:surf_model}, the Kapton thickness is found to be \edit11{24.74$\pm$0.04~$\mu$m} and the Be thickness is \edit1{14.20$\pm$0.01~$\mu$m} with a reduced $\chi^2$ error of 2.48. These effective thicknesses derived from the Henke model are typically different from the physical thickness, but are used self-consistently between calibration and subsequent analysis.

From the model fit to the NIST SURF spectra in Figure~\ref{fig:surf_model} the effective area of DAXSS is found from Equation~\ref{eqn:resp} without the $G(E_{det},E_{ph})$ term.
\begin{figure}[H]
    \centering
    \includegraphics[width=\linewidth]{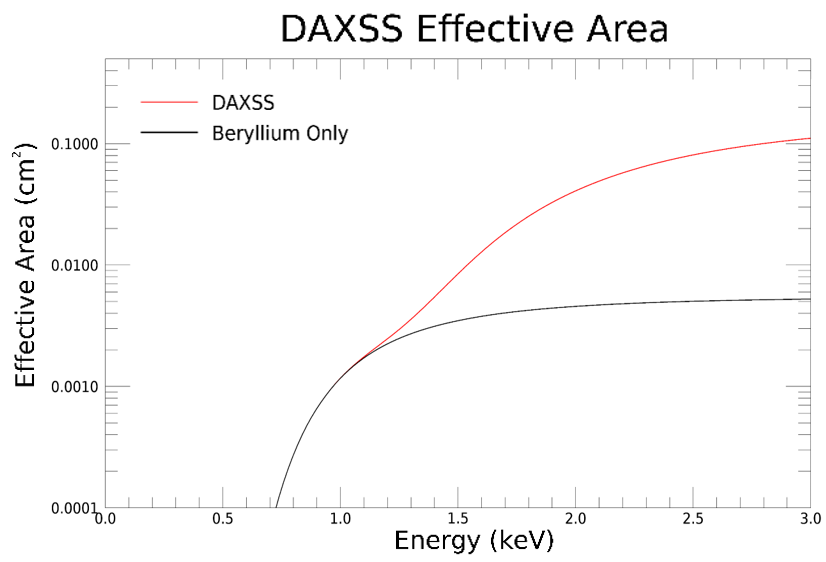}
    \caption{\label{fig:eff_area} Effective area from the Be contribution compared to that of the full Kapton and Be filter for the DAXSS instrument. Here it is clear to see how the Kapton filter aperture enhances the transmission of photons $\gtrapprox$1.2~keV.}
\end{figure}
Using the effective area in Figure~\ref{fig:eff_area} and the photon redistribution described by a Gaussian with FWHM in Equation~\ref{eqn:fano} the \edit1{DRA} from Equation~\ref{eqn:resp} can be determined. This \edit1{DRA} can now be used with DAXSS solar measurements to estimate the true solar fluxes during observations. This process is described for the 2018 June rocket flight in the following section.  

\section{Observation and Spectral Model Fits}\label{sec:model}
\begin{figure}[H]
    \centering
    \includegraphics[width=\linewidth]{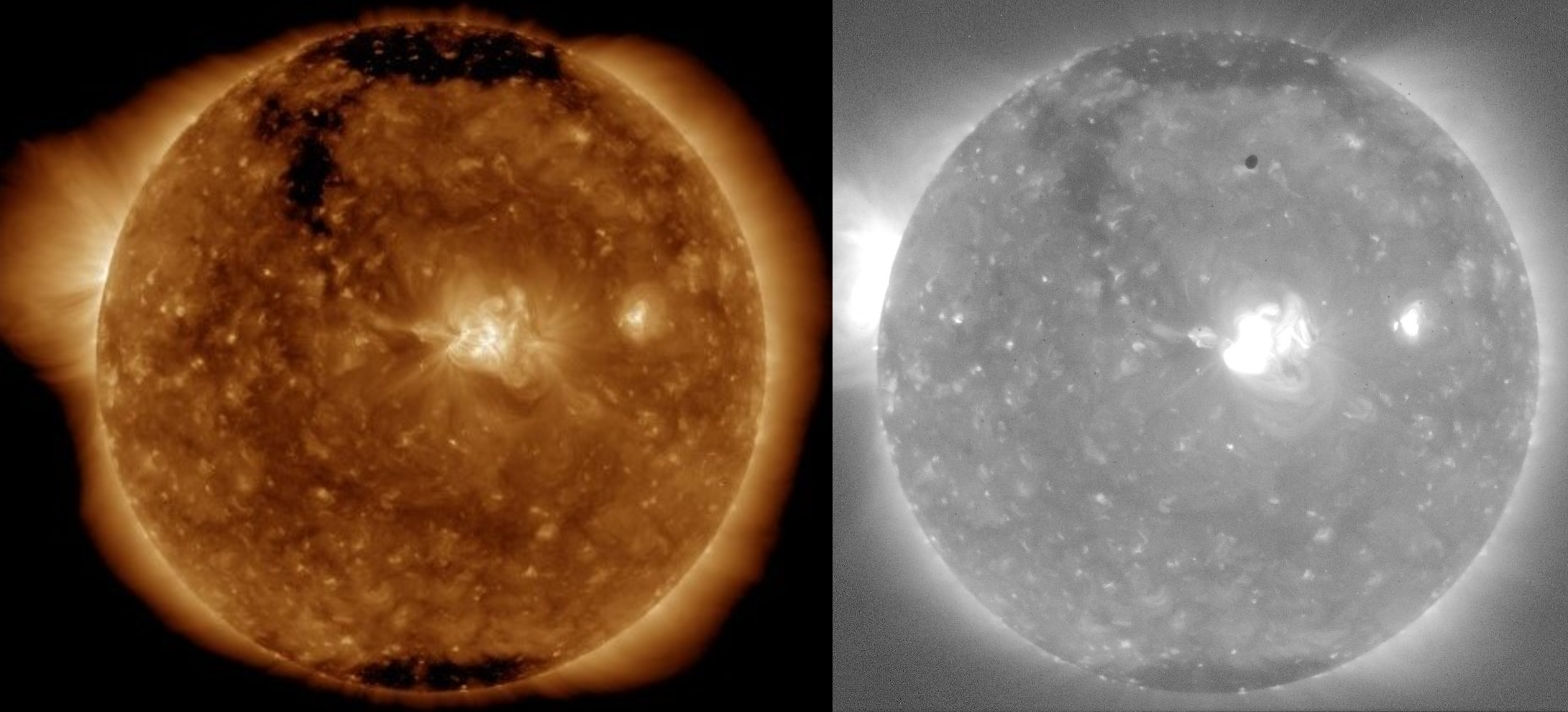}
    \caption{SDO AIA 193~{\AA} and Hinode XRT images during the 2018 June 18 rocket launch. Two small active regions can be seen near the middle of the solar disk and another behind the east limb. With the absence of large active regions and solar flares, the sun can be classified as being in a quiescent state.  \label{fig:images}}
\end{figure}

The DAXSS measurements analyzed here were taken on 2018 June~18 at approximately 19:05~UT over a span of roughly 3.5~minutes. \edit1{This observation was during a time with quiescent (non-flaring) small active regions on the solar disk and when the 10.7 cm radio flux (F10.7) was 75 solar flux units (1 sfu = 10–22 W/m$^2$/Hz).} Figure~\ref{fig:images} shows an SDO Atmospheric Imaging Assembly (AIA) and Hinode X-ray Telescope (XRT) images from the day of launch. 

The CHIANTI atomic database, v9.0.1 \citep{dere_1997,dere_2019}, is used to provide model spectra of solar photon intensity versus energy. This model includes both the continuum and emission lines for given user inputs of solar temperature, emission measure, abundance, and ionization fractions. The default CHIANTI ionization fractions were used for all of the CHIANTI calculations expressed in this paper. Following the process by \citet{caspi2015}, CHIANTI spectra are generated using the IDL SolarSoft function \texttt{f\textunderscore vth.pro}. This function returns a model solar radiation spectrum as seen at Earth in units of photon flux (photon cm$^{-2}$ s$^{-1}$ keV$^{-1}$). To convert from this output spectrum to detector count rate (counts s$^{-1}$) the output model spectrum is multiplied by the \edit1{detector} response array \edit1{(DRA)} \edit1{outlined in Sec.~\ref{detector_responsivity}} (units of cm$^{2}$ keV counts photon$^{-1}$).
 
%%\begin{figure}
%%    \epsscale{1.2}
%%   \plotone{figures_pdf/AIA_DEM_2018-06-18}
%%    \caption{AIA differential emission measure for 2018 June 18, a few hours after the launch. The active region in the center of the sun has a higher emission measure than the quieter surrounding solar disk in the temperature ranges $log(T)$ = [6.00,6.40], or 1--2.5~MK, and $log(T)$ = [6.40,6.80], or 2.5--6.3~MK.}
%%    \label{fig:AIA_DEM}
%%\end{figure}

The model spectrum was smoothed to match the energy resolution of DAXSS \edit1{so that} a direct comparison \edit1{can be} made to the measured spectrum. The smoothing function that most closely replicates the X-123 detector is a normalized Gaussian function with a full width half maximum (FWHM) that is variable with energy as described in Figure~\ref{fig:res}. The model spectrum is smoothed by convolving with the Gaussian of variable FWHM. The smoothed model irradiance spectrum \edit1{is multiplied by the detector response array to convert to a model in count space and can be compared directly to the detector's measured spectrum in count space since they are} in the same units and compatible in energy resolution.

The goodness of fit is characterized quantitatively by finding the reduced $\chi^2$ of the model with respect to the measured spectra. The count rate error, or variance array, used for each energy bin in the $\chi^2$ calculation assumes standard Poisson counting statistics and is found by taking the square root of the counts in each bin, divided by the integration time.

The fitting region is set to span from 0.7--\edit1{3.5}~keV for all of the following model fits. The lower limit is set due to the need for a better detector response model for energies below 0.7~keV as this is where instrumental effects become dominant over the solar signal due to the increasing attenuation of the filter (see Figure~\ref{fig:eff_area}). This is attributed to excess detector counts due to photon energy-loss processes where higher-energy photons lose some energy before being detected, and hence count as low-energy photons. These processes can include photoelectron emission in the Be window, Si-K escape, Si-L escape, and Compton scattering as theoretically modeled by \citep{moore2018}, but the specific contributions of these processes have not yet been verified experimentally. The upper limit is due to a low amount of detected photons with energies above \edit1{3.5}~keV which causes a larger amount of photon-counting uncertainty. \edit1{The value of 3.5~keV was selected because the total counts accumulated over the measurement duration was less than 10 for energies higher than 3.5~keV.}

In this paper we describe a model that calculates a best fit using one\edit1{temperature (1T)} or two temperatures \edit1{(2T)} with respective emission measures in addition to a singular \edit1{AF}, which affects only \edit1{low-FIP} elements, or a relative AF that varies for each element with emission lines observable above the continuum. The singular AF is a constant that is multiplied by the Feldman standard extended coronal (FSEC) abundance values \citep{feldman1992,Landi2002}. The relative AF for each identifiable element is a multiplication factor to the FSEC abundance value for that element while keeping the values for other elements unchanged.

\edit1{For each model fitting routine a Monte Carlo method of 1,000 iterations is used for both selecting the in initial guess for the parameters as well as adding counting statistical noise to each energy bin. Each parameter is multiplied by a different randomly selected factor from a uniform distribution between 0.5 and 2. This was repeated for all iterations giving unique initial starting parameters each time. The measured count rate of each energy bin is altered by adding its countrate uncertainty multiplied by a different randomly selected factor from a normal distribution. This was also repeated over all iterations, which gave a unique model spectra each time.}

\edit1{It is important to note the all standard deviations reported in Tables~\ref{table:1T_model}, \ref{table:2t_sing}, and \ref{table:2t_abun} are the standard deviations of the mean value found by 1,000 model fits and not the actual uncertainties of the physical values. The uncertainties quoted by MPFIT are almost always way too small as well. This is in part because it assumes the uncertainties are uncorrelated, when they are actually highly correlated and the T, EM, and AF parameters are not completely independent of each other. For best-fit temperature values found in each model a fair physical temperature uncertainty is about 0.05~MK and the best-fit EM and AF values have about a 10($\%$) uncertainty in their physical values stated.}

\subsection{1T and 2T Models with Single Abundance Factor}
\begin{figure*}[!h]
    \centering
    \includegraphics[width=\textwidth]{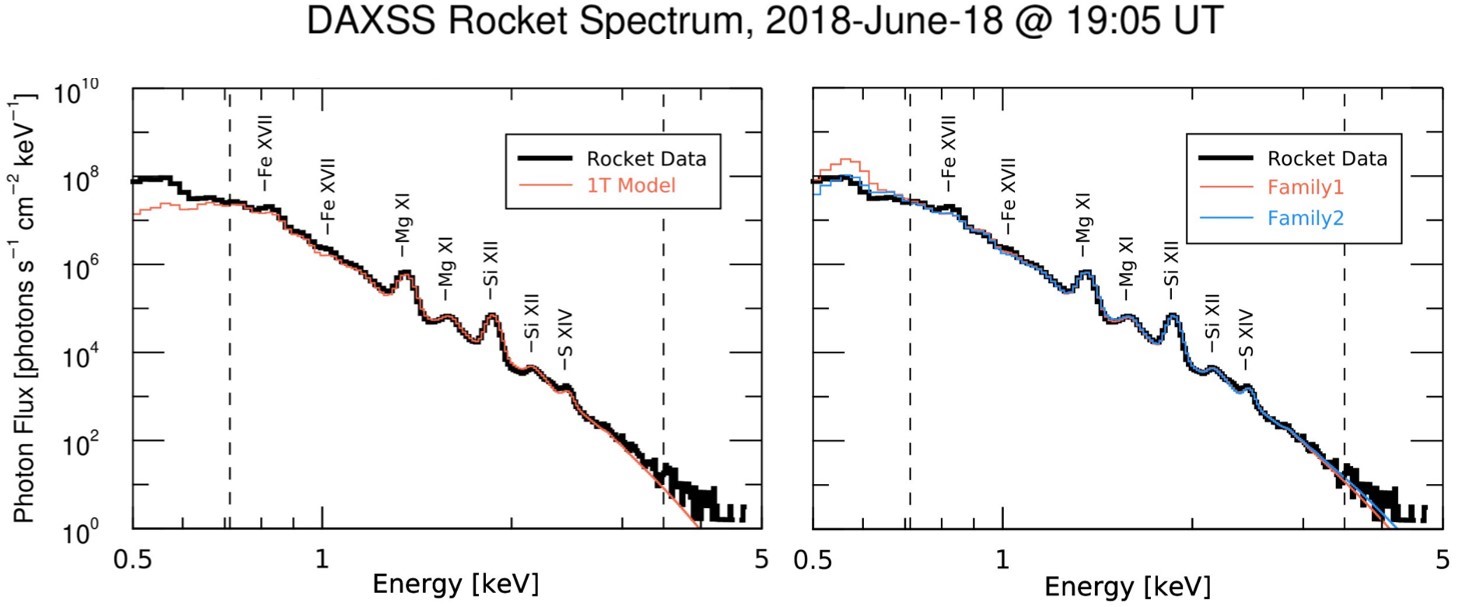}
    \caption{\label{fig:1T_2T}1T and 2T models with singular AF. Fit range is between 0.7 and 3.5~keV (vertical dashed lines). (Left) 1T model with single temperature, single emission measure, and singular AF. The best-fit parameter values are shown in Table~\ref{table:1T_model}. The 1T model is the most simplistic but does not fit well over the whole energy range. (Right) 2T single AF model fit of the flight averaged DAXSS spectrum shows good agreement with the measured data and fits the whole energy range better than a single-temperature model fit. The best-fit parameters for both families are displayed in Table~\ref{table:2t_sing}}
\end{figure*}

The first fit was made assuming the corona was a uniform, \edit1{isothermal} plasma \edit1{described by} a single temperature, single emission measure, and single abundance factor. The parameters were adjusted by an IDL fitting procedure \texttt{mpfit}. This fitting procedure iteratively changes the parameters to find a minimum $\chi^2$ value using the Levenberg-Marquardt technique. \edit1{The best-fit parameter values are shown in Table~\ref{table:1T_model} and the best-fit model spectrum is shown in Figure~\ref{fig:1T_2T} (Left). The reduced chi square value for the 1T model was found to be 7.7.}

\begin{deluxetable*}{l c c c c}[!h]
\tablecaption{1T Fit with Singular AF\label{table:1T_model}}
\tabletypesize{\scriptsize}
\tablehead{\colhead{} & \colhead{T} & \colhead{EM} & \colhead{AF} & \colhead{Red.Chi-Sq} \\
 \colhead{} & \colhead{(MK)} & \colhead{($\num{e49}\ cm^{-3}$)} & \colhead{($\times$ FSEC)} & \colhead{}
} 
\startdata
    {Mean} & 2.85 & 0.129 & 0.91 & 7.7 \\ 
    {Stdev} & 0.01 & 0.002 & 0.01 & 0.2 
\enddata
\tablecomments{This is a single temperature and single emission measure fit with a singular abundance factor. T represents temperature, EM represents emission measure, and AF represents a singular abundance factor multiplied by Feldman Standard Extended Coronal (FSEC) values \citep{feldman1992}.}
\end{deluxetable*}

\begin{figure*}[!h]
    \centering
    \includegraphics[width=\textwidth]{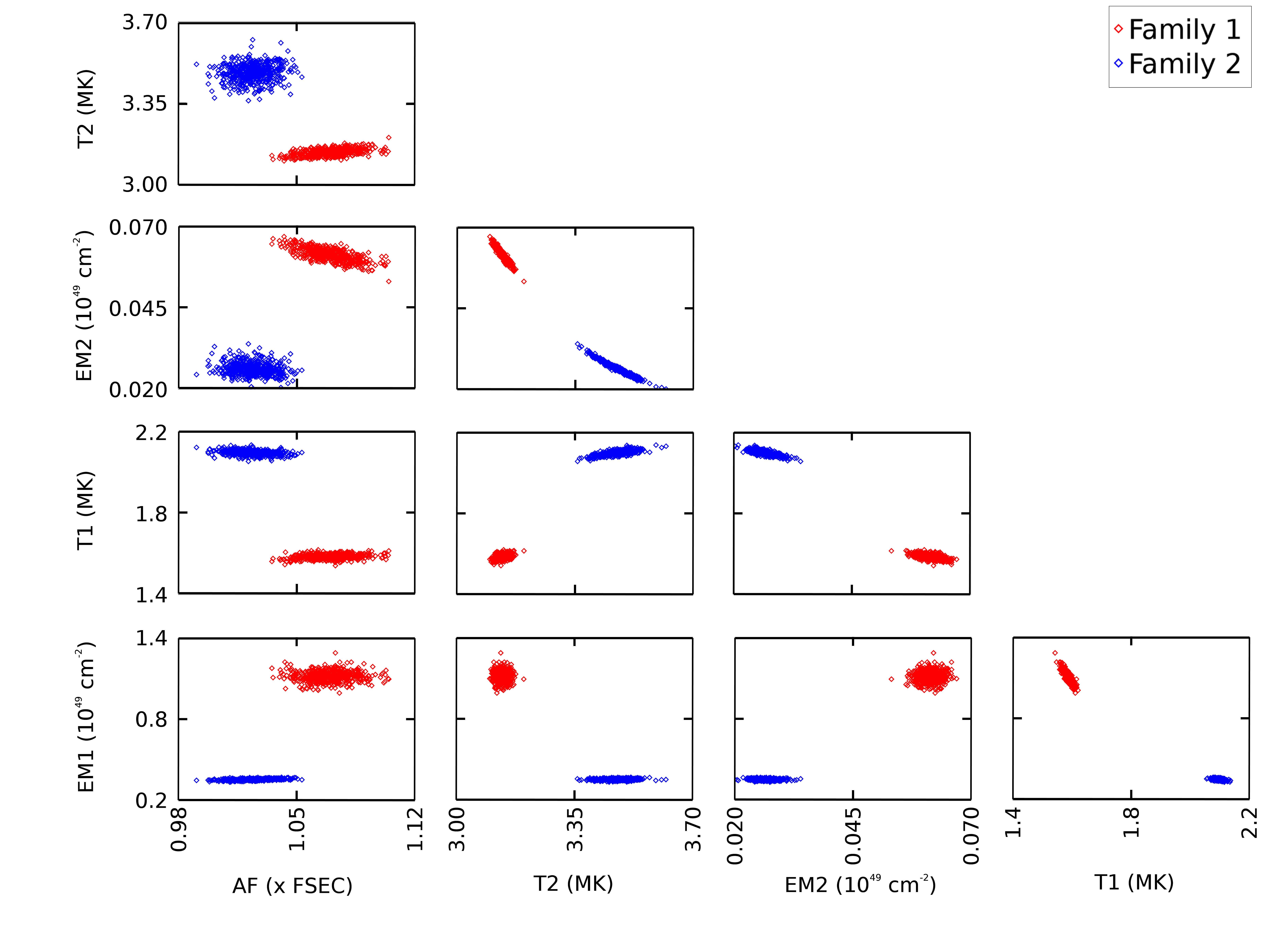}
    \caption{\edit1{2T single AF fit parameter correlations from 1,000 Monte Carlo runs of fitting model. The family groups are evident in this plot and do not overlap in any parameter correlation subplot. They are considered separate when computing each mean parameter value. Each family has similar $\chi^2$ values so both are regarded as valid fits in the 2T single AF model that approximates the sun as only two temperatures and two emission measures with a single AF.}}
    \label{fig:2T_param_corr}
\end{figure*}

The \edit1{1T model} starts to fall away from the measured data at energies above 2.5~keV. This indicates that there is some coronal plasma at a higher temperature. Therefore, a second fit approach was done to allow for solar plasma at two different temperatures. The same \edit1{energy range was used for fitting} as before, 0.7--\edit1{3.5}~keV. This \edit1{2T single AF} fit has five free parameters to be adjusted: two temperatures, two emission measures, and a single abundance factor. \edit1{Out of the 1,000 fits that were found, some of the fit values arranged in a normal distribution about one mean and others arranged in a normal distribution about a separate mean with their standard deviations not overlapping. The reason for this is that the 2T single AF approximation is sampling the true DEM at two points, and the ``families" are simply different sample points to which the fit converges. Although the models produced by each family have} similar $\chi^2$ \edit1{values, meaning their}  goodness of fit \edit1{was comparable, they could not be considered of the same family and such could not be averaged together to find the mean value for each parameter. Justification for this is shown in the parameter correlation plot of Figure~\ref{fig:2T_param_corr}.} 

\edit1{The best-fit parameter values of both families had to be averaged separate, and are listed in the Table~\ref{table:2t_sing}.} Figure~\ref{fig:1T_2T} (Right) shows the model produced by each family of best-fit values. The \edit1{2T single AF} fit follows the measured spectra more closely into higher energies while maintaining a good fit at the lower energies as well. The reduced $\chi^2$ value using the \edit1{2T single AF} fitting method decreased to \edit1{4.5 for family 1 and 4.9 for family 2,} indicating a closer model fit to the DAXSS measurement. There is a visible disagreement between the modeled and measured spectra at around 0.82~keV. The explanations to this discrepancy are discussed in the conclusion (Sec. \ref{sec:conclusion}).

\subsection{2T Multiple AF Model} \label{subsec:2t_rel_abund}
The parameters found in the \edit1{1T} fits for temperature and emission measure can be passed back into CHIANTI to produce a line list of ions contributing to spectrum. This line list allows us to calculate the estimated elemental emission contribution to each of the blended lines we see in the rocket measurement, as shown in Table \ref{table:cont}. From this it is evident that the line features seen in the DAXSS spectra are dominated by Mg, Si, S, and Fe using this CHIANTI line list.

\begin{figure}[!h]
    \centering
    \includegraphics[width=\linewidth, trim = {35 210 70 230}, clip]{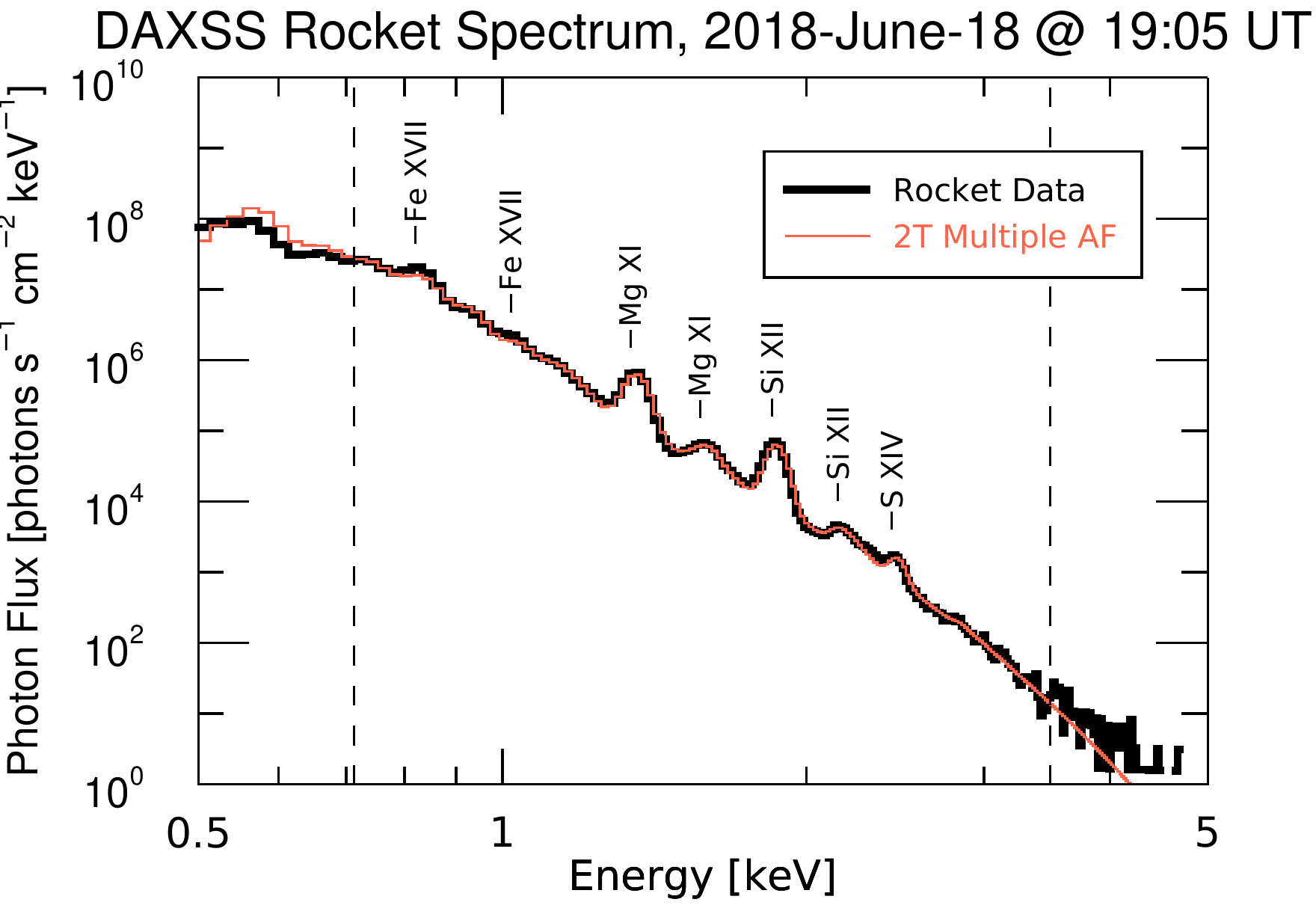}
    \caption{\label{fig:fit_2t_rel} 2T multiple AF model fit with variable parameter abundances of four identifiable low-FIP elements: Mg, Si, S, and Fe. Fit range is between 0.7 and 3.5~keV (vertical dashed lines). This 2T multiple AF model shows the best agreement with the measured DAXSS data. The best-fit parameter values are reported in Table~\ref{table:2t_abun}.}
\end{figure}

As such, in a third fit, shown in Figure~\ref{fig:fit_2t_rel}, the relative abundances of the four low-FIP elements Mg, Si, S, and Fe are made to be free parameters in addition to two temperatures and their respective emission measures. The addition of individual abundance factors for each identifiable element in the DAXSS spectrum can account for the abundances in the corona deviating from FSEC abundance values. One thousand iterations of this fitting process was also done \edit1{and this time} only a single family of parameters was found. The relative abundances that give the lowest $\chi^2$ fit and therefore the best fit are listed in Table~\ref{table:2t_abun}.

\begin{deluxetable*}{l c c c c c c c}[!h]
\tablecaption{DAXSS Solar SXR Model Measurements with Singular Abundance Factor\label{table:2t_sing}}
\tabletypesize{\scriptsize}
\tablehead{\colhead{} & \colhead{T1} & \colhead{EM1} & \colhead{T2} & \colhead{EM2} & \colhead{AF} & \colhead{Red.Chi-Sq} & \colhead{Percentage}\\
 \colhead{} & \colhead{(MK)} & \colhead{($\num{e49}\ cm^{-3}$)} & 
\colhead{(MK)} & \colhead{($\num{e49}\ cm^-3$)} & \colhead{($\times$ FSEC)} & \colhead{} & \colhead{($\%$)}
} 
\startdata
    {Family 1 Mean} & 1.59 & 1.11 & 3.14 & 0.061 & 1.07 & 4.5 & 48.9 \\ 
    {Family 1 Stdev} & 0.01 & 0.04 & 0.02 & 0.002 & 0.01 & 0.2 & \nodata \\
     & & & & & & & \\
    {Family 2 Mean} & 2.10 & 0.35 & 3.48 & 0.026 & 1.02 & 4.9 & 50.3\\ 
    {Family 2 Stdev} & 0.01 & 0.01 & 0.04 & 0.002 & 0.01 & 0.2 & \nodata\\
\enddata
\tablecomments{This is a two temperature and two emission measure fit with a singular abundance factor. T represents temperature, EM represents emission measure, and AF represents a singular abundance factor multiplied by Feldman Standard Extended Coronal (FSEC) values \citep{feldman1992}. The percentage column represents the number of iterations out of 1,000 that showed up in each family.}
\end{deluxetable*}

\begin{deluxetable*}{c c c c}[!h]
\tablecaption{Leading Elemental and Ion Emission Contributions to Peak Features in 2018 Jun. 18 DAXSS Spectra.
\label{table:cont}}
\tabletypesize{\scriptsize}
\tablehead{
\colhead{Peak Location} & \colhead{Primary} & \colhead{Secondary} & 
 \\ \colhead{(keV)} & \colhead{(\%)} & \colhead{(\%)} 
} 
\startdata
    0.81 & {Fe: 90.19 (XVII, XVI)} & {O: 9.11 (VIII)}\\
    0.91 & {Ne: 88.42 (IX, VII)} & {Fe: 8.16 (XVII)} \\
    1.01 & {Fe: 63.67 (XVII)} & {Ne: 27.30 (X, IX)}\\
    1.35 & {Mg: 98.83 (XI, X)} & {\nodata} \\
    1.57 & {Mg: 66.96 (XI, X)} & {Al: 32.01 (XII)}\\
    1.85 & {Si: 98.83 (XII, XIII)} & {Al: 1.12 (XI)}\\
    2.13 & {Si: 99.42 (XII, XIII)} & {\nodata} \\
    2.43 & {S: 99.97 (XIV, XV)} & {\nodata} \\
\enddata
\tablecomments{CHIANTI spectral parameters of $T_1=2.10\ MK$ and $T_2=3.48\ MK$ with $EM_1=\num{0.35e49}\ cm^{-3}$ and $EM_2=\num{0.026e49}\ cm^{-3}$ with FSEC abundance. Elements only listed if contributing $\geq$1\%.}
\end{deluxetable*}

\begin{deluxetable*}{l c c c c c c c c c}[!h]
\tablecaption{DAXSS Solar SXR Model Measurements with Abundance Factor for Identifiable Elements\label{table:2t_abun}}
\tabletypesize{\scriptsize}
\tablehead{
\colhead{} & \colhead{T1} & 
\colhead{EM1} & \colhead{T2} & 
\colhead{EM2} & \colhead{Mg AF} & \colhead{Si AF} & \colhead{S AF} & \colhead{Fe AF} & \colhead{Red.Chi-Sq} \\
 \colhead{} & \colhead{(MK)} & \colhead{($\num{e49}\ cm^{-3}$)} & 
\colhead{(MK)} & \colhead{($\num{e49}\ cm^-3$)} & \colhead{($\times$ FSEC)} & \colhead{($\times$ FSEC)} & \colhead{($\times$ FSEC)} & \colhead{($\times$ FSEC)} & \colhead{}
} 
\startdata
    Mean & 1.86 & 0.50 & 3.29 & 0.043 & 1.02 & 0.99 & 1.00 & 1.35 & 3.9 \\ 
    Stdev & 0.02 & 0.02 & 0.03 & 0.003 & 0.01 & 0.02 & 0.05 & 0.02 & 0.2 \\
\enddata
\tablecomments{These values are from a two temperature and two emission measure fit with a separate abundance factor for each identifiable element in the DAXSS spectrum. T represents temperature, EM represents emission measure, and AF represents an abundance factor for each element multiplied by its respective Feldman Standard Extended Coronal (FSEC) value \citep{feldman1992}. All other FSEC abundance values are maintained as their original value.}
\end{deluxetable*}

The relative \edit1{AF} for Mg, Si, and S \edit1{are almost unity showing that they} are all within 2 percent of their FSEC abundance values. The relative \edit1{AF} of Fe, however, was found to be \edit1{35} percent higher than its FSEC abundance value. When adding more free parameters to a model it is expected that the reduced $\chi^2$ will decrease, but there is also a risk that the model strays further from physical reality. This discrepancy and feature is further discussed in the conclusion (Sec. \ref{sec:conclusion}).

\section{Comparisons}\label{sec:compare}
\subsection{Previous SXR Measurements}
\begin{figure}[!h]
    \centering
    \includegraphics[width=\linewidth]{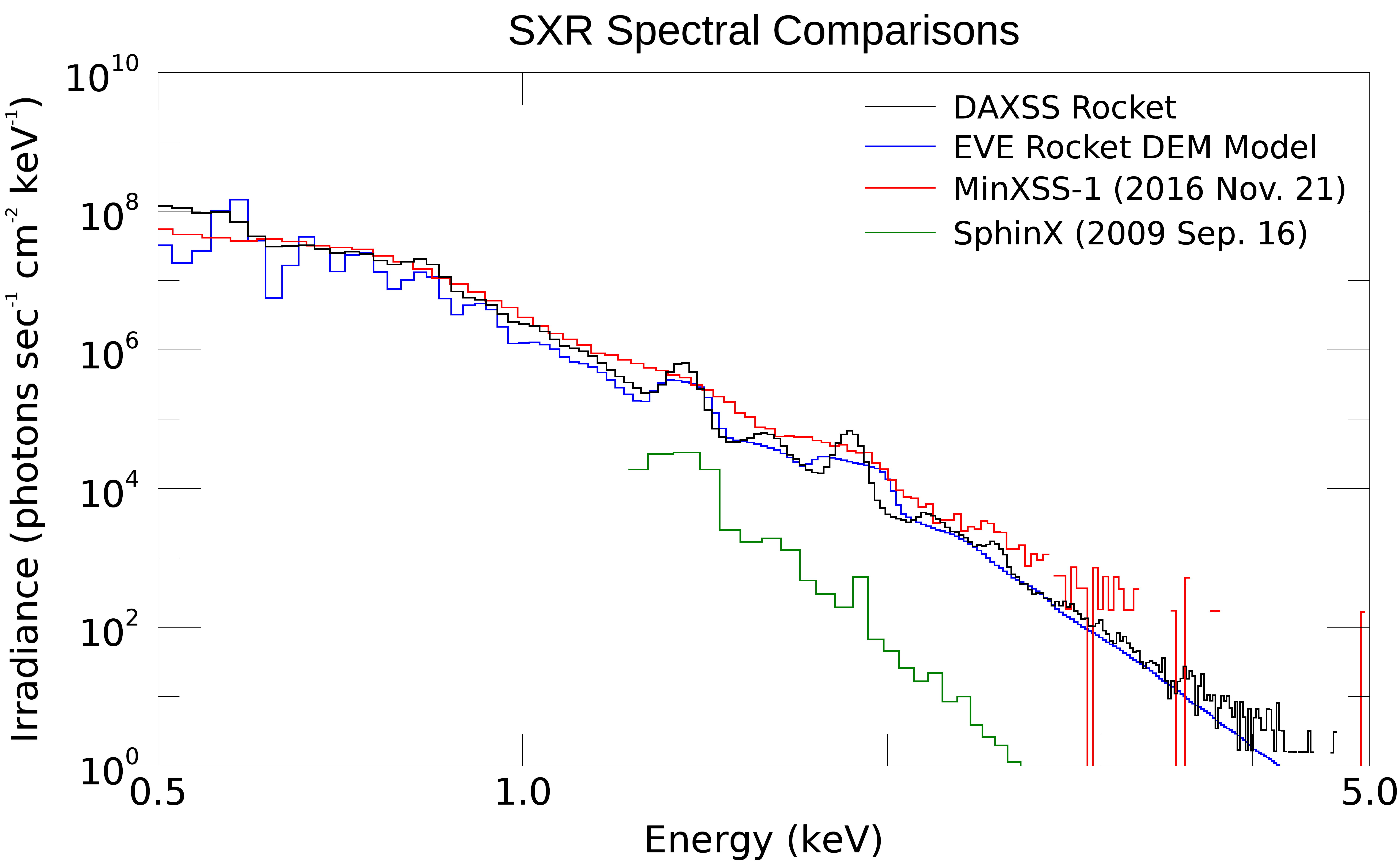}
    \caption{Comparisons from other Si-diode spectrometer measurements to  the 2018 June 18 DAXSS rocket flight (solar activity levels and spectral fitting for each spectra shown in Table \ref{table:comp}).  The EVE spectrum is created by deriving a DEM from the rocket EVE's Multiple Extreme ultraviolet Grating Spectrograph (MEGS-A: 6-37 nm) spectral observation during the rocket flight (Subsec. \ref{subsec:dem}), and this DEM is then used in CHIANTI to generate a SXR spectrum at the resolution of DAXSS.\label{fig:sxr_comp}}
\end{figure}

The 2018 June 18 DAXSS rocket measurement can be validated by other solar SXR observations. From Figure~\ref{fig:sxr_comp} one can see that MinXSS-1 measurements, from similar solar conditions, and derived SXR spectra from simultaneous EVE MEGS-A EUV measurements have very similar spectra as the DAXSS measurement. Additionally, from Table \ref{table:comp} one can see that \edit1{the 1T and 2T} models, derived during similar solar conditions, yield similar results to those described in Sec.\ref{sec:model}, more so for the lower temperature component. Notably, the SphinX measurement at lower solar activity has a cooler temperature and lower emission measure than the DAXSS measurement. Furthermore, there are hotter temperatures for the more active measurements in 2003 and 2012.
 
From Table \ref{table:res_comp} and Figure~\ref{fig:sxr_comp} it is shown that DAXSS offers significantly improved energy resolution in the SXR spectrum compared to the previous photon-counting Si-diode SXR spectrometers and thus can better spectrally resolve and identify some elemental line features. As shown in Figure~\ref{fig:res}, the resolution varies with energy, and the DAXSS resolution is about three times better than MinXSS-1 at the lower energy (1~keV) and about 50\% better at higher energy (10~keV).  Crystal spectrometers, such as the Bragg Crystal Spectrometer (BCS) on Yohkoh \citep{yohkoh_bcs_1991} and  CORONAS/RESIK \citep{resik_1998} and the Bent Crystal Spectrometer on the Solar Maximum Mission (SMM) \citep{smm_1980}, offer much higher spectral resolution than seen on Si-diode spectrometers and as such can offer absolute abundance measurements over their narrow passbands. For example, the Yohkoh BCS observes the S XV emission over 5.02 -- 5.11~{\AA} (2.5~keV) and the Ca XIX emission over 3.16 -- 3.20~{\AA} (3.9~keV) \citep{bcs_yohkoh1996}. However, as DAXSS has a wider passband, and improved resolution from previous Si-diode instruments, it has the benefit of making simultaneous inferred abundance measurements from multiple elements, as well as temperature and emission measure measurements (as done in Subsec. \ref{subsec:2t_rel_abund}). Furthermore, measurements from these previous crystal spectrometers did not observe elemental lines below $\sim$2.4~keV, meaning that DAXSS can provide improved insight into abundance, temperature, and emission measure models for the SXR between 0.5 -- 2.5~keV.

 \subsection{DEM Comparisons}\label{subsec:dem}
\begin{figure}[H]
    \centering
    \includegraphics[width=\linewidth]{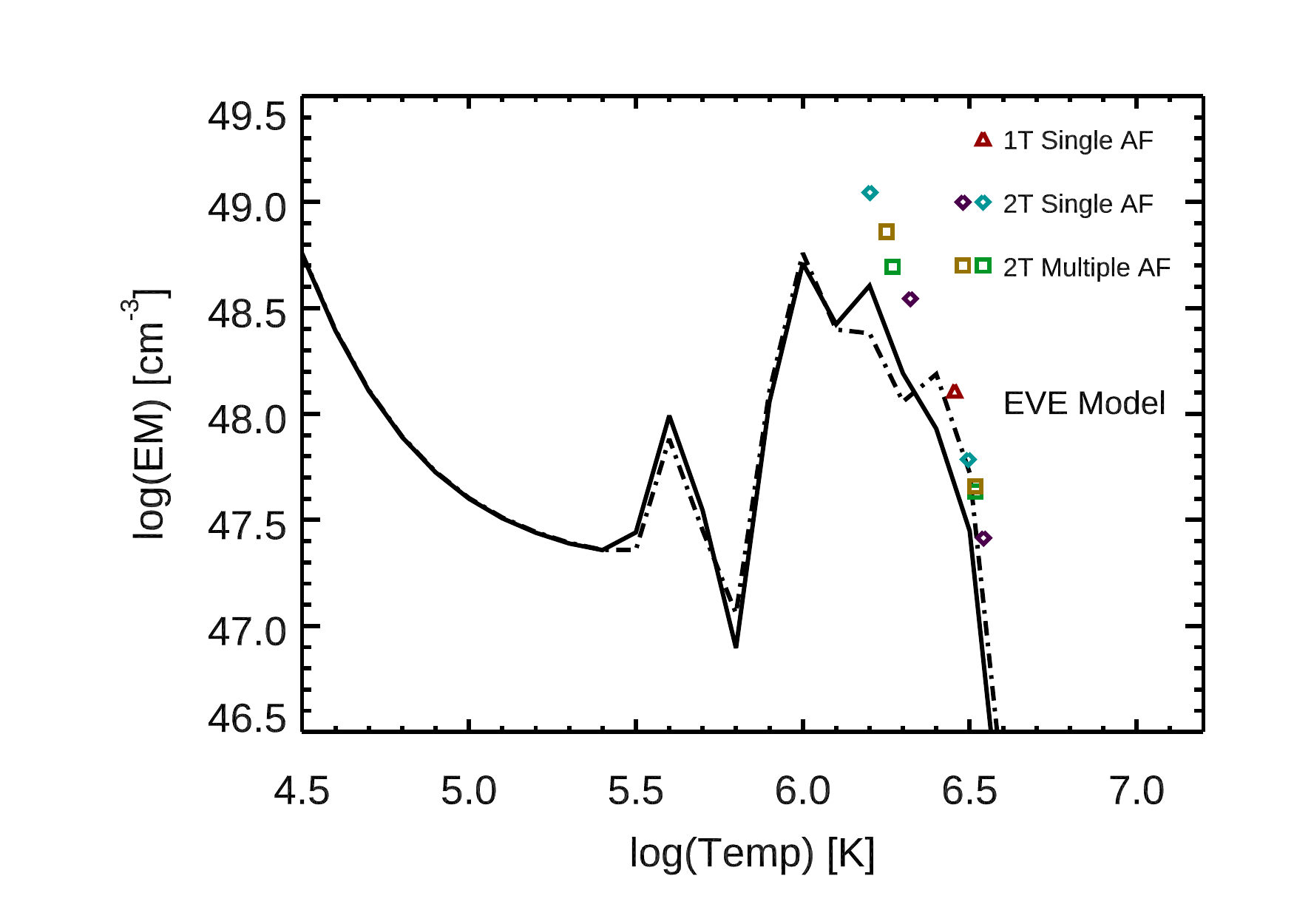}
    \caption{Comparison of DAXSS 1T Single AF model (Table~\ref{table:1T_model}, triangle), 2T Single AF model (Table \ref{table:2t_sing}, diamonds) and 2T multiple AF model (Table \ref{table:2t_abun}, squares) to rocket EVE-derived EM results (black solid line). The EVE-based DEM model with QS and AR DEMs is the dot-dashed line.}
    \label{fig:EM_comp}
\end{figure}

 Comparison of the DAXSS two-temperature (2T) model fits to differential emission measure (DEM) estimates from other instruments could help validate the DAXSS simple modeling approach.  \citet{moore2018} shows that the X123 response is primarily over the temperature range of 1 MK to 10 MK, so this comparison needs to be done over a similar range, such as is accessible with using DEMs derived from extreme ultraviolet (EUV) emissions. One of the comparison is with the DEM derived with Solar Dynamics Observatory (SDO) \edit1{\citep{pesnell_2012}} Extreme ultraviolet Variability Experiment (EVE) \edit1{\citep{woods_2012,hock_2012}} solar EUV spectral irradiance data in the 6~nm to 37~nm range.  The derivation of EVE-based DEM estimates is being developed for improving the X-ray ultraviolet Photometer System (XPS)  data processing for the Solar Radiation and Climate Experiment (SORCE) \edit1{\citep{woods_2005a}} and Thermosphere, Ionosphere, Mesosphere, Energetics, and Dynamics (TIMED) \edit1{\citep{woods_2005b}} missions.  The primary DEMs for this analysis is the quiet sun (QS) and active region (AR) DEMs as needed for estimating the daily variations of the solar XUV spectral irradiance for the XPS Level~4 product \citep{woods_2008}.  The \ion{Fe}{8} to \ion{Fe}{16} lines in the SDO EVE spectra were initially used to estimate the DEMs for the reference QS and AR spectra derived with EVE data between 2010 and 2013 using the technique described by \citet{Schonfeld_2017}.  It was then found that fitting with DEM Gaussian profiles with logarithmic Temperature (K, log(T)) peaks every 0.2 and with Gaussian width of 0.42 in log(T) (FWHM of 1.0) provided more robust solutions for the DEM (similar technique described by \citet{warren_2013}). Furthermore, fitting just specific Fe lines was providing low irradiance estimates in the 6~nm to 15~nm range.  Better spectral model values for this range was found when fitting the EVE spectra over the ranges of 10 to 14~nm and 26 to 30~nm.  The DEM estimate using just the rocket EVE spectral data flown with DAXSS on 2018 June 18 and the DEM based on combining the QS DEM and the AR DEM with an AR scaling factor of 0.00806 are shown in Figure~\ref{fig:EM_comp}. This scaling factor for the AR EM was determined as the best fit for the DAXSS spectral irradiance. The conversion of the EVE-based DEM ($cm^{-5} K^{-1}$) to EM ($cm^{-3}$) for comparison to DAXSS 2T \edit1{model} solution is the multiplication by the solar hemisphere area ($\num{3.04E22} cm^2$) and by the temperature bin size (0.23 * Temperature in K). 
 
 The DAXSS family of 2T \edit1{model} solutions for the SXR range have very similar emission measures (EM) profile over temperature as \edit1{the DEM profile} derived from the EVE data in the EUV range, but the DAXSS \edit1{2T model} EM values are a factor of \edit1{2.8} higher. Most of this difference is \edit1{expected} because the DAXSS 2T solution is only at two temperatures whereas the EVE result is a DEM over many temperatures.  The difference between the EVE model \edit1{D}EM and an iso-thermal EM \edit1{near 2 MK} is estimated to be about a factor of 4. However, some of this difference can be attributed to the DAXSS 2T \edit1{model} solutions having a lower abundance factor than the \citet{feldman1992} abundance values used in CHIANTI for the EVE-based DEM modeling \citep{feldman1992}. The 10 percent decrease in the abundance factor for the DAXSS 2T solution does increase its EM relative to an EM result using the standard (unadjusted) \citet{feldman1992} abundance values used in CHIANTI for the EVE-based DEM modeling. It is most important to note that the DAXSS 2T model results with multiple solutions (families) appear to all be valid solutions because they follow the same profile over temperature as the EVE-based DEM model results.

\subsection{Validation for GOES XRS}
The rocket DAXSS calibrated SXR spectral irradiance also provides a validation for the new generation of the X-Ray Sensor \(XRS\) aboard the GOES-16 and GOES-17 operational satellites \citep{goes2020}. \edit1{For a meaningful comparison, the GOES XRS processing algorithm needs to be described because XRS is a broadband measurement whose irradiance result greatly depends on what is assumed for the solar spectrum. Equations 7-8 provide the irradiance conversion for XRS to its Level 2 product, \(E_{L2}\). The pre-flight calibrations include the aperture area, A in units of \( \si{\meter\squared}\), and responsivity, R in units of \( \si{\ampere\per\watt}\). An assumed reference solar spectrum, \(E_{ref}\) in units of \si{\watt\per\meter\squared\per\nm}, is also required for the irradiance conversion. The standard GOES XRS processing } assumes a flat spectrum, that is, no wavelength variation for \(E_{ref}\). As listed in Table \ref{table:goes_comp}, the GOES-16 and GOES-17 solar irradiance Level 2, \(E_{L2}\), values for the {1-8 \AA}  XRS-B channel are \num{6.18E-8} \si{\watt\per\meter\squared} and \num{6.67E-8} \si{\watt\per\meter\squared}\, respectively \@. \edit1{This is only an estimate of the SXR irradiance because the assumption for a flat spectrum is not realistic.}

\edit1{The use of a flat spectrum introduces a large irradiance bias that needs to be removed before doing a comparsion to DAXSS. Equations 9-10 prescribe how to compare the SXR spectral measurements from DAXSS to the broadband measurement of XRS. The accurate comparison is between the measured current from XRS, \(I_{measure}\), and the predicted current with DAXSS spectra, \(I_{predict}\). The true XRS irradiance, \(E_{XRS_{true}}\), is then calculated by Equation 10.} The DAXSS irradiance integrated over the XRS-B {1-8 \AA}  band is \num{4.59E-8} \si{\watt\per\meter\squared}\@. \edit1{The DAXSS spectrum is used as the reference} solar spectrum \edit1{along with the XRS} instrument spectral responsivity, R \edit1{, in Equation 9 to } provide a predicted XRS \edit1{sensor current} of {0.81 pA} and {0.83 pA} for GOES-16 and 17, respectively.  Then comparison of these predicted signals to the measured XRS signals provides more accurate irradiance results (Equation 10) from GOES-16 and 17 XRS-B of \num{5.17E-8} \si{\watt\per\meter\squared} and \num{5.45E-8} \si{\watt\per\meter\squared}, respectively. These are 13\% and 19\% higher than the DAXSS {1-8 \AA} irradiance. 

\begin{equation}
    I_{measure}=E_{L2}\cdot R_{integ}
\end{equation}
\begin{equation}
    R_{integ}=\frac{\int_0^\infty R(\lambda)\cdot A \cdot E_{ref}(\lambda)\cdot d\lambda}{\int_{\lambda_1}^{\lambda_2}E_{ref}(\lambda)\cdot d\lambda}
\end{equation}
\begin{equation}
    I_{predict}=\int_0^\infty R(\lambda)\cdot A \cdot E_{DAXSS}(\lambda)\cdot d\lambda
\end{equation}
\begin{equation}
    E_{XRS_{true}} = \frac{I_{measure}}{I_{predict}}\cdot E_{DAXSS}
\end{equation}

\begin{deluxetable*}{l c c c c c c}[!t]
\tablecaption{Solar temperature and emission measure comparisons during quiet or non-flaring solar activity.\label{table:comp}}
\tabletypesize{\scriptsize}
\tablehead{
\colhead{Observation} &\colhead{P10 Index} & \colhead{Observatory} & \colhead{Temperature 1} & 
\colhead{Emission Measure 1} & \colhead{Temperature 2} & 
\colhead{Emission Measure 2} \\
 \colhead{(YYYY-MM-DD)} & \colhead{(s.f.u.)} & \colhead{} &
\colhead{($\log$ MK)} & \colhead{($\num{e49}\ cm^{-3}$)} & 
\colhead{($\log$ MK)} & \colhead{($\num{e49}\ cm^{-3}$)} 
} 
\startdata
    {2018-06-18} & {75.14} & {NASA 36.336 (DAXSS)$^\alpha$} & {1.56} & {1.13} & {3.08} & {0.068} \\ 
    {2016-11-21 -- 2016-11-22} & {75.19} & {MinXSS-1$^\beta$} & {2.36} & {0.28} & {5.00} & {0.008} \\
    \begin{tabular}[c]{@{}l@{}}{2012-06-23}\\{\citep{caspi2015}}\end{tabular} & {112.91} & {NASA 36.286} & {2.9} & {0.49} & {11} & {0.0014}\\
    \begin{tabular}[c]{@{}l@{}}{2009-09-16}\\{\citep{sylwester2012}}\end{tabular} & {70.28} & {SphinX} & {1.71} & {0.0978} & {\nodata} & {\nodata}\\
    \begin{tabular}[c]{@{}l@{}}{2003}\\{\citep{resik2010}}\end{tabular} & {\nodata} & {RESIK$^{\gamma}$} & {2.9} & {0.17} & {9.1} & {0.023}\\
\enddata
\tablecomments{$^\alpha$DAXSS measurement from Family 1 in Table~\ref{table:2t_sing}. $^\beta$MinXSS-1 measurements are determined by fitting two temperatures and EMs over the spectrum from 0.9--2.7~keV. $^\gamma$Averaged data taken from early 2003 when GOES levels were between A9 -- B1.  P10 index is defined as the average of the F10.7 cm radio flux and 80-day smoothed F10.7 cm radio flux.   }
\end{deluxetable*}

\begin{deluxetable*}{l c c }[!h]
\tablecaption{Resolution Comparison of Wide Passband Spectrometers\label{table:res_comp}}
\tabletypesize{\scriptsize}
\tablehead{
\colhead{Instrument} &\colhead{Resolution (keV)} & \colhead{Energy (keV)} 
} 
\startdata
    {DAXSS} & {0.069 -- 0.228} & {0.7 -- 20} \\
    {MinXSS-1} & {0.214 -- 0.341} & {1.0 -- 30} \\
    \begin{tabular}[c]{@{}l@{}}{CORONAS-PHOTON/SphinX}\\{\citep{sylwester2012}}\end{tabular}& {0.464$^\alpha$} & {1.2 -- 14.9} \\
    \begin{tabular}[c]{@{}l@{}}{MESSENGER/SAX}\\{\citep{sax2015}}\end{tabular} & {0.6} & {6}\\
    \begin{tabular}[c]{@{}l@{}}{GSAT-2/SOXS}\\{\citep{soxs_2006}}\end{tabular} & {0.7} & {6}
\enddata
\tablecomments{$^\alpha$SphinX resolution quoted as the same over its entire passband.}
\end{deluxetable*}

\begin{deluxetable*}{l c c}[!b]
\tablecaption{GOES XRS comparison to DAXSS\label{table:goes_comp}}
\tabletypesize{\scriptsize}
\tablehead{
\colhead{} & \colhead{GOES-16} & \colhead{GOES-17} \\ 
\colhead{} &\colhead{XRS-B: {1-8 \AA}} &\colhead{XRS-B: {1-8 \AA}} }
\startdata
    \(E_{L2}:\) NOAA-reported XRS L2 Irradiance (W/m$^2$) & $\num{6.18e-8}\ \pm\ \num{0.3e-8}$ & $\num{6.67e-8}\ \pm\ \num{0.4e-8}$ \\
    \(R_{integ}:\) XRS Response Integrated (A/(W/m$^2$)) & $\num{9.615e-6}$ & $\num{9.577e-6}$ \\
    \(I_{measure}:\) XRS Sensor Current Measured (A) & $\num{9.07e-13}$ & $\num{9.86e-13}$  \\
    \(I_{predict}:\) XRS Sensor Current Predicted (A) & $\num{8.05e-13}$ & $\num{8.30e-13}$ \\
    DAXSS $1-8 \AA$ Irradiance (W/m$^2$) & $\num{4.59e-8}\ \pm\ \num{0.46e-8}$ & $\num{4.59e-8}\ \pm\ \num{0.46e-8}$ \\
    \(E_{XRS_{true}} :\) XRS "True" Irradiance (W/m$^2$) & $\num{5.17e-8}\ \pm\ \num{0.8e-8}$ & $\num{5.45e-8}\ \pm\ \num{0.8e-8}$ \\
    Ratio (DAXSS Irrad./GOES "True" Irrad.) & $0.89\ \pm\ 0.16$ & $0.84\ \pm\ 0.16$ 
\enddata
\end{deluxetable*}

  Considering about 10\% accuracy for each instrument responsivity as DAXSS and GOES XRS instruments were each calibrated at SURF using similar techniques, the 1-sigma uncertainty for this comparison is estimated to be 18\%. While the XRS difference to DAXSS is close to this 1-sigma uncertainty, we suspect that the XRS signals could still have a particle (energetic electrons) background signal contribution, which is a known concern for the GOES solar observations in its GEO orbit. The GOES-16 and 17 XRSs have been cross-calibrated to agree at higher flare irradiance levels, so the 5\% difference between GOES-16 and GOES-17 for this comparison at low solar activity is also suggestive that their corrections for particle background signals could be improved.  The flight of DAXSS-2 in 2021 on the EVE calibration rocket will provide additional validation for the GOES XRS.  If this 2021 flight is during a period of higher solar activity (as expected for solar cycle 25), then we anticipate a more accurate comparison from having larger signals (higher signal-to-noise), and also a validation for the XRS-A {0.5-4 \AA} channel could be possible.
 
\section{Conclusions}\label{sec:conclusion}
The DAXSS instrument with the Amptek X-123 FAST SDD spectrometer provides improved measuring capabilities in the soft X-ray regime than previous flights of X123-type technology. The latest rocket spectrum has better energy resolution by more than a factor of two and improved sensitivity over a wider energy range than its previous flights because of the improvements that Amptek has made to the X-123 and due to the dual-zone aperture design for DAXSS. These improvements allow DAXSS to better fill a long-standing gap in the SXR region of solar observations with more detailed information. The measurements from DAXSS show lines with improved spectral resolution, compared to prior Si photodiode measurements, allowing for improved coronal temperature and emission measure models, as well as the ability to more accurately fit abundances per individual element.

A two-temperature (2T) model \edit1{with both a single AF and multiple AF} was fit to the measured rocket DAXSS spectrum and \edit1{fit the measured spectrum better, with} a lower $\chi^2$ value, than a single temperature \edit1{(1T) model}. \edit1{Both of the} 2T model\edit1{s} agree with the measured data from lower photon energies of 0.7~keV out to higher energies of \edit1{3.5}~keV. \edit1{Both of} the 2T model temperatures and emission measures are also consistent with the EVE-derived DEM profiles.

The 2T model fit included both a fit for a singular abundance factor and with multiple abundance factors for some select elements. For \edit1{the single AF approach}, the resulting AF was \edit1{$1.07 \pm\ 0.01$ for family one and $1.02 \pm\ 0.01$ for family two. For the multiple AF approach the resulting AF was $1.02 \pm\ 0.01$ for Mg, $0.99 \pm\ 0.02$ for Si, $1.00 \pm\ 0.05$ for S, and a much higher than expected $1.35 \pm\ 0.02$ for Fe, as} seen in Table~\ref{table:2t_abun}. \edit1{It is of interest that} the abundance of iron was found to be \edit1{35} percent higher than the currently accepted FSEC abundance value for \edit1{a non-flaring,} quiescent sun. \edit1{The DAXSS instrument takes whole-sun measurements and such cannot determine exactly where the emission it measures is coming from. Referring to Figure~\ref{fig:images} we can see that there were two active regions visible near the center of the solar disk over the duration of data accumulation. Although the quiet sun emits SXR radiation, active regions emit the vast majority of the measured intensity since quiet sun irradiance levels are much lower when active regions are not present \citep{sylwester2012}. This does, however, not explain the higher abundance of Fe that was reported.} 

\edit1{The are many possible reasons that could explain the higher abundance of Fe.}The model that we are using, which only includes two temperatures and two emission measures, could be insufficient for modeling the solar spectrum and may not accurately represent the temperature contributions in this region of the spectrum. This would lead to incorrect line-to-continuum ratios and improper abundance values. Creating a DEM model based on the DAXSS spectrum might clear up this possibility. Another explanation for this large discrepancy in the abundance of iron could be that there are missing elemental emission lines in the CHIANTI line database near those energies, which then are not represented in its modeled solar spectrum. It could also be that there is a higher abundance of the element emitting at 0.82~keV than the FSEC abundances, although this would go against many prior measurements including in the EUV and X-rays. To resolve the discrepancy whether the excess flux is a legitimate measurement or due to a simplistic model, or the need for updated spectral fitting or atomic modeling, we need additional high-resolution, high-sensitivity long-term systematic measurements from instruments in this energy range in the future.

There is on-going work to improve the instrument response model at energies below 0.7~keV. At those lower energies, the X-123 response is affected greatly by the photoelectric effect on its Be filter, Si-K escape, Si-L escape, and Compton scattering causing diversion away from the simplified response model based directly on the NIST SURF calibrations. Future differential emission measure (DEM) analysis is also planned in order to obtain more complete temperature coverage, and will include SDO and Hinode observations to expand the temperature coverage even further.

Sounding rocket flights are a great platform to demonstrate new technology, such as the DAXSS instrument, as well as to investigate the properties of the sun on a particular day. Daily observations over several years are needed to achieve \edit1{better understanding} for how the coronal heating processes, plasma temperature, and composition change for active regions during different solar cycle phases, and higher cadence observations of the order of a few seconds are needed for flare energetics studies. The rocket DAXSS instrument is being prepared for a flight of opportunity on the InspireSat-1 with its launch planned for late 2020. This flight will provide science data needed by solar physicists to study solar active region evolution and flare energetics and to provide evidence that discriminates between competing models of coronal heating through detailed analysis of the solar SXR spectra.

\edit1{\subsection{Acknowledgements}}
\edit1{This research was performed with funding from the NASA Grant NNX17AI71G to the University of Colorado (CU). We deeply thank the SDO EVE instrument team, the NASA sounding rocket program and its NASA Sounding Rocket Operations Contract (NSROC) staff, and the White Sands Missile Range staff for their outstanding dedication and support for the flight of DAXSS on the SDO EVE calibration rocket payload in June 2018. We are also immensely thankful for the NIST SURF staff for their excellent support of DAXSS calibrations at their facilities in Gaithersburg, MD. The majority of this research was done by CU graduate students Bennet Schwab and Robert Sewell, and they are vastly grateful to CU Professors Scott Palo and Heather Lewandowski for their graduate academic advice and scientific guidance.  }

\bibliography{DAXSS_Rocket_Paper_2019}{}
\bibliographystyle{aasjournal}

\end{document}